\documentclass[conference,a4paper]{IEEEtran}
\usepackage{xcolor}
\usepackage{balance}
\usepackage{cite}

\ifCLASSINFOpdf
  \usepackage[pdftex]{graphicx}
\else
  \usepackage[dvips]{graphicx}
\fi

\usepackage{calc,ifthen,intcalc}
\usepackage{tikz}
	\usetikzlibrary{arrows,shapes,positioning,calc,fadings,patterns,decorations}
	\usetikzlibrary{decorations.pathreplacing,decorations.pathmorphing}			
	\usetikzlibrary{arrows,arrows.meta}
	\usetikzlibrary{shapes.geometric}

\usepackage{amsmath}
\usepackage{bm,amssymb}

\ifCLASSOPTIONcompsoc
 \usepackage[caption=false,font=normalsize,labelfont=sf,textfont=sf]{subfig}
\else
 \usepackage[caption=false,font=footnotesize]{subfig}
\fi

\usepackage{lipsum}
\usepackage{booktabs}

\newcommand\figref[1]{Fig.~\ref{#1}}
\newcommand\tabref[1]{Tab.~\ref{#1}}

\begin{document}

\title{Enhancing IoT Network Reliability through Antenna Diversity in Real-World Scenarios}

\author{\IEEEauthorblockN{
        Marco Niederberger\IEEEauthorrefmark{1},
        Yanick Schoch\IEEEauthorrefmark{1},
        Michel A. Nyffenegger\IEEEauthorrefmark{1},
        Christian Gwerder\IEEEauthorrefmark{2},
        and Hans-Dieter Lang\IEEEauthorrefmark{1}
    }
    %\\
    \IEEEauthorblockA{\IEEEauthorrefmark{1}
        OST Eastern Switzerland University of Applied Sciences, Rapperswil, SG, Switzerland, Email: hansdieter.lang@ost.ch}
    \IEEEauthorblockA{\IEEEauthorrefmark{2}
        Steinel Solutions AG, Einsiedeln, Switzerland}
}

\maketitle

\begin{abstract}
    Antenna diversity in Internet of Things (IoT) devices is explored to address challenges associated with fixed device orientation and interference from nearby metallic objects, which can degrade wireless network performance.
    Integrating multiple antennas with different polarizations can significantly enhance signal quality and network reliability.
    Experiments demonstrated significant improvements in link quality, with gains exceeding 30\,dB in received signal strength for specific scenarios. Already the addition of a second PCB-based antenna alone, resulted in a 17\,dB improvement.
    These findings suggest that the added complexity in hardware and software to incorporate antenna diversity could well be justified by the resulting enhancements in network stability for real-world applications.
\end{abstract}

\vskip0.5\baselineskip
\begin{IEEEkeywords}
    IoT, antenna diversity, wireless communication, network reliability, received signal strength.
\end{IEEEkeywords}

\vspace*{2pt}
\section{Introduction}
The Internet of Things (IoT) has rapidly expanded into various facets of everyday life.
Besides enhancing the convenience of various daily activities, it has also started revolutionizing industries and intersections of these areas.
IoT devices are found in smart homes, industrial automation, healthcare, and urban infrastructure, among other applications, enabling everything from simple monitoring and control to real-time data collection, analysis, and decision-making \cite{rose15,asghari19}.

Certain critical tasks, for example healthcare monitoring and industrial automation, require not only a broad range of capabilities but also highly reliable and efficient network connectivity to ensure functionality, safety, and data integrity.

Evaluating and improving wireless communication reliability has been the subject of a lot of research,
with antenna diversity, being one of the prominent hardware-oriented strategies \cite{tradeoffStudy,lee72,yoo17}. Previous studies have explored theoretical aspects of antenna diversity, such as the mathematical modeling of signal propagation and polarization effects \cite{golmohamadi16,chowdhury22,razmhosseini20}. Some work has also been done on the hardware design of multi-antenna systems, often focusing on optimizing the size and cost of antennas for practical deployment \cite{ramirez17,ramirez18}.
In contrast, this paper explores the practical potential of incorporating multiple simple antennas with different polarizations into the devices,
to enhance the performance and reliability of wireless links within the IoT network in real-world scenarios and confirm these with measurements.

\vspace*{4pt}
\section{Problem Statement}
In many applications and specific environments, the position and orientation of each IoT device are often constrained and cannot be freely chosen. Moreover, once these devices are installed, altering their location or orientation is typically not possible or would incur significant costs. Additionally, installations are often carried out without proper consideration of wireless signal propagation. Lastly, even if these considerations are initially addressed, changes in the environment over time can introduce new complications at a later point.

\begin{figure}[!htb]\centering
    \vspace*{-3.5mm}
    \subfloat[][]{\includegraphics[width=3.8cm]{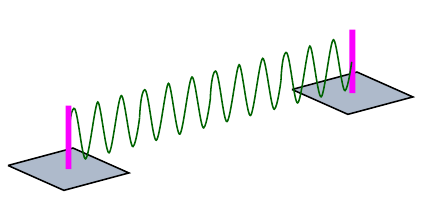}}
    \hfill
    \subfloat[][]{\includegraphics[width=3.8cm]{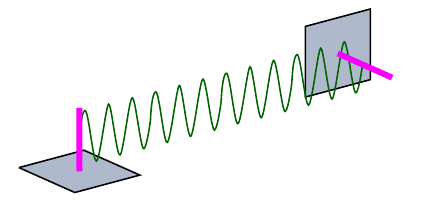}}
    \caption{Impact of antenna alignment on link quality: (a) optimal alignment, and (b) poor connection due to misalignment (polarization mismatch).}\label{fig:misalignment}
    \vspace*{0.5mm}
\end{figure}

All of these conditions present significant challenges for establishing a reliable wireless network. Most importantly, after the installation, it may turn out that the antennas are badly misaligned, as illustrated in \figref{fig:misalignment}.
Additionally, adverse effects from nearby objects, particularly metallic constructions, commonly occur.
In environments with high variability, such as factories, warehouses, or public spaces, changes like moving people, large equipment, robots, or even installations can affect the network.
These shifting conditions can drastically degrade the quality, reliability, and throughput of connections in a simple wireless system that lacks flexibility.

Therefore, IoT system providers are questioning whether simple solutions could mitigate these issues if they arise.
Particularly compelling is the idea of including multiple antennas and dynamically selecting the optimal one during operation\,---\,without being overly intrusive or changing the overall communication scheme.
One approach involves adding just one additional antenna, resulting in a two-antenna setup with a switch to select between them.
This raises the question of whether two orthogonal PCB antennas in the same plane already provide sufficient means to address the aforementioned challenges,
or if a third antenna is necessary to offer full freedom in terms of polarization.
While exploring different antennas with varying radiation patterns could also provide valuable insights, this aspect is beyond the scope of this study.
Additionally, implementing full MIMO communication, which uses multiple antennas simultaneously, is not yet feasible for short-range IoT mesh networks, due to the lack of availability of compatible radio chips and standards support \cite{IEEE:standard}.
\newpage

\section{Related Work}

While the general concept of antenna diversity, particularly polarization diversity, is well-known in mobile communications \cite{yoo17,lee72}, the literature review revealed relatively limited practical exploration of this topic for stationary IoT mesh networks.
Nonetheless, several studies and vendor white papers have investigated various approaches to use antenna diversity to improve IoT network reliability.

An application note by Silicon Labs \cite{silicon15} tested an approach where the best antenna was selected during the protocol's preamble and used for the payload in both outdoor open spaces and indoor multipath environments.
Outdoors, minimal improvement was noted due to a lack of reflections.
Indoors, where reflections and movement significantly affected the performance, nodes without antenna diversity required a 6\,dB increase in transmit power to achieve comparable Packet Error Rates (PER).
These results demonstrate the effectiveness of their approach for indoor connections.

Golmohamadi and Frolik \cite{golmohamadi16} conducted experiments by mounting two antennas first in an anechoic chamber and then in a reverberation chamber to measure the path loss for different combinations of polarization.
Both the transmitter and the receiver were positioned in all three orthogonal directions.
In the anechoic chamber, the difference in received power between co-polarized and cross-polarized (polarization mismatched) antennas reached up to 22\,dB.
In the reverberation chamber, the difference was reduced to around 15\,dB, due to the presence of multiple paths between the transmitter and receiver, which mitigated the impact of polarization mismatches.

Ramirez et\,al. \cite{ramirez17} showed that intelligent switching between three orthogonal polarizations could enhance connection quality by up to 11\,dB.
They used additive manufacturing to create three quarter-wavelength monopoles on an ABS substrate in a mutually orthogonal configuration.
In our study, antennas are integrated onto a flat PCB due to design constraints.
Further work by Ramirez et\,al. \cite{ramirez18} explored different configurations using a Wilkinson power combiner, achieving a median improvement of 4.3\,dB.
This work highlighted the system's sensitivity to slight variations in antenna placement, approximately $\lambda$/10, which significantly affected link quality.

A recent white paper by Qorvo \cite{qorvo24} introduced ``ConcurrentConnect" technology, using a hardware-based Rx diversity algorithm to enhance communication range and reliability in indoor environments.
Leveraging the IEEE 802.15.4 standard \cite{IEEE:standard} and SNR correlation for antenna selection, Qorvo's method improved the communication link budget by approximately 7 to 12\,dB, depending on the setup.
In contrast to this study's focus on mesh networks, Qorvo's focus was on point-to-point connections, demonstrating different strengths and potential applications for antenna diversity in IoT devices.

Lastly, \cite{chowdhury22} and \cite{razmhosseini20} also investigate antenna diversity for multihop IoT mesh networks.
However, they focus more on theoretical analysis and broader communication metrics, such as capacity and signal correlation, making their results difficult to compare to this practical study.

\vspace*{4pt}
\section{Methodology}

\subsection{Hardware Design}
\subsubsection{RF Components}
\begin{figure}\centering
    \includegraphics[width=.99\columnwidth,clip,trim={11mm 2mm 11mm 16mm}]{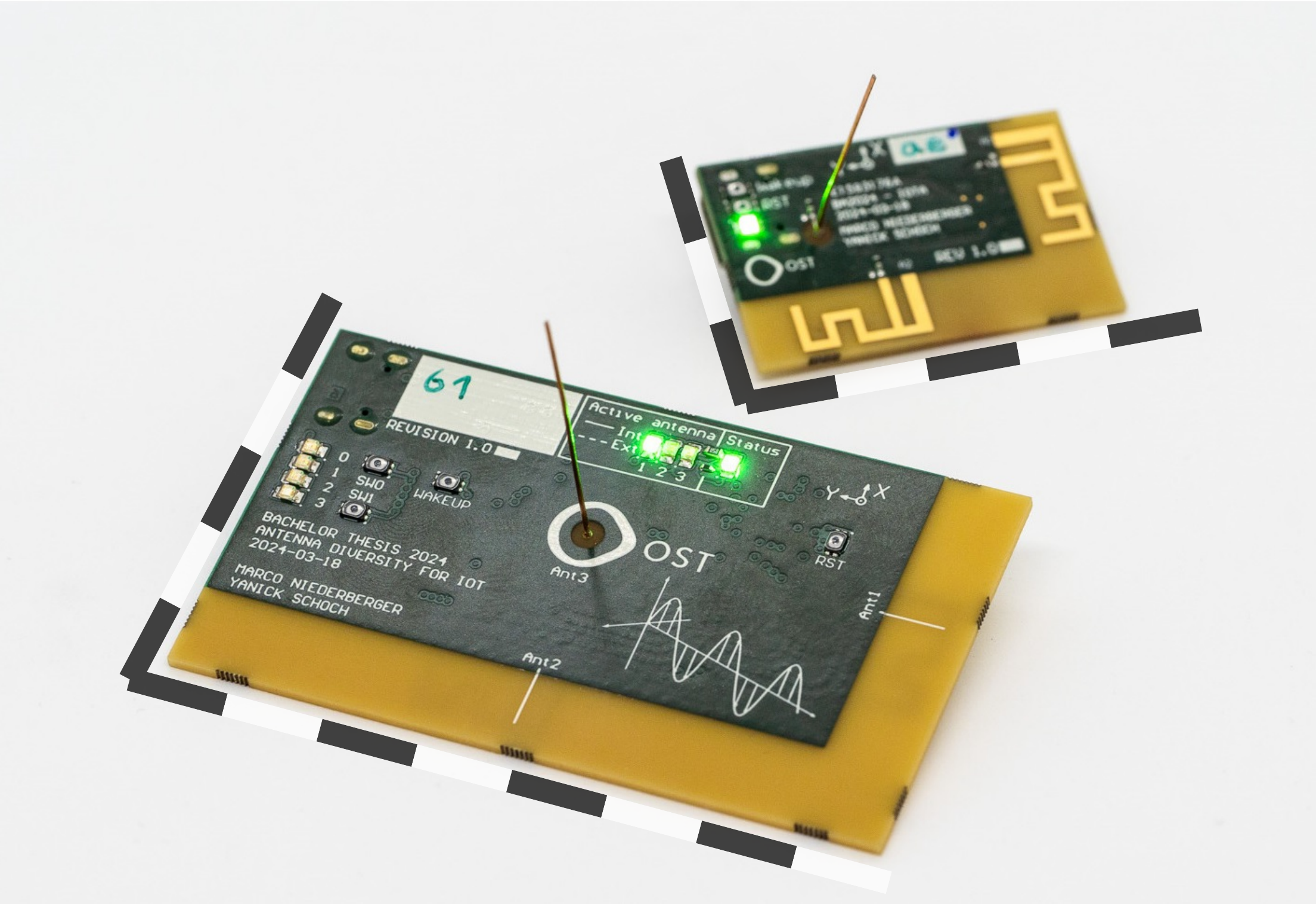}
    \caption{The two developed IoT nodes: larger (74$\times$43\,mm, front) and smaller (40$\times$25\,mm, back), each equipped with three orthogonal antennas\,---\,two IFAs on the PCBs and a wire monopole. The rulers show centimeter intervals.}
    \label{fig:nodes}
\end{figure}

For the following investigations, two types of IoT nodes were developed: a larger and a smaller one, as shown in \figref{fig:nodes}. Each node is equipped with the Nordic nRF\,52840 low-power multiprotocol SoC \cite{nordic22} and a Qorvo QPC8014Q single-pole 6-throw (SP6T) switch \cite{qorvo16}. This switch is notable for its low insertion loss of only 0.33\,dB within the operational frequency band around 2.44\,GHz.

\subsubsection{Antennas}
Each node features three orthogonally oriented antennas: two PCB-based inverted-F antennas and a centrally positioned wire monopole antenna, intentionally kept simple to maintain a straightforward design.  For practical purposes, note that the inclusion of the third, wire monopole antenna adds significant challenges, particularly in managing its physical integration, as it protrudes from the otherwise two-dimensional PCB layout.
All antennas are tuned to the ISM frequency band of 2.4 to 2.48 GHz, achieving a return loss of 10\,dB or better, with coupling figures below --11\,dB for the large node and --8\,dB for the small node.

\subsubsection{Sensors and Battery}
The nodes are constructed with a compact layout that balances the need for multiple antennas with the constraints of small size and power efficiency, typical for IoT devices. Integrated accelerometers and magnetometers enable the system to dynamically select the optimal antenna based on the node's orientation and surrounding environmental conditions. This capability allows for real-time optimization of network performance, crucial for maintaining robust communication links in diverse and often unpredictable deployment scenarios.
Additionally, both node types are equipped with a battery management controller and rechargeable LiPo batteries, to support untethered operation in the field.

\subsection{Software Design}
For this study, the mesh network is based on OpenThread \cite{OT}, an open-source implementation of the Thread protocol.
Thread is an IPv6-based wireless networking protocol designed for low-power IoT applications and provides robust mesh networking capabilities \cite{Thread}. The OpenThread network connects to the local intranet via a Raspberry Pi acting as an OpenThread BorderRouter.

\begin{figure}[!htb]\centering
    \includegraphics[width=0.95\columnwidth]{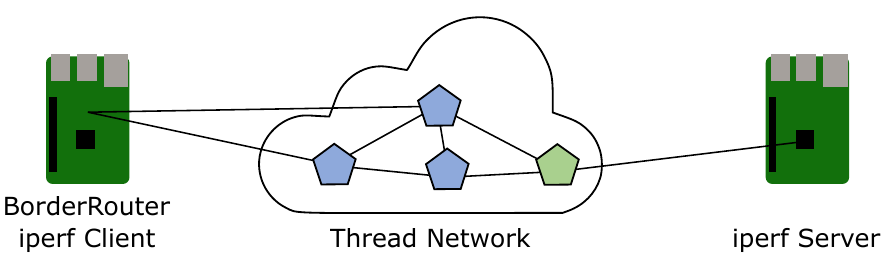}
    \caption{Throughput measurements between two Raspberry Pis at both ends of the network.
        In this study, the Raspberry Pis were positioned in close proximity to the final node to avoid bottlenecks in the last network link.}
    \label{fig:throughput_setup}
\end{figure}

This Raspberry Pi is also used as iPerf client, as shown in \figref{fig:throughput_setup}, facilitating an accurate assessment of the network's data transmission capabilities.
iPerf is a widely-used network testing tool that provides active measurements of the maximum achievable bandwidth and the achieved package error rate on IP networks.
It is highly versatile, supporting various protocols and enabling detailed testing scenarios.

By generating data streams and measuring the effective data rate at which they are transferred across the network, iPerf provides valuable insights into the performance characteristics.
Beyond assessing the impact of different antenna configurations for this antenna diversity study, also potential bottlenecks could be identified.

Finally, a browser-based visualization tool was developed to provide enhanced control over the test setup, enabling real-time interaction with the network and performance monitoring. This tool allows users to select the active antenna for each node, monitor RSSI values for all connections, and visualize the network layout\,---\,all within a single web interface.

\subsection{Experimental Procedure}
\subsubsection{Test Environment}
\begin{figure}[!htb]\centering
    \subfloat[][]{\includegraphics[width=0.99\columnwidth,clip,trim={0 33mm 0mm 0}]{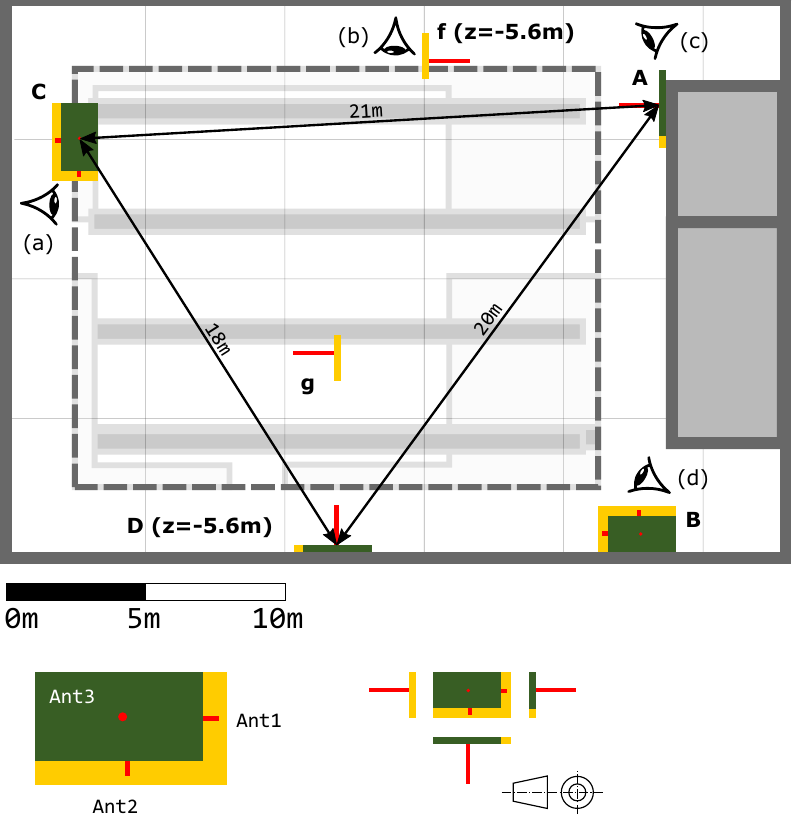}}

    \vspace*{-1mm}
    \subfloat[][]{\includegraphics[width=4.4cm,clip,trim={0 0mm 79mm 110mm}]{paper6_plan_embedded.pdf}}
    \hfill
    \subfloat[][]{\includegraphics[width=4.4cm,clip,trim={55mm 0mm 29mm 110mm}]{paper6_plan_embedded.pdf}}

    \caption{Test network built in a two-story building: (a) shows the positions of all nodes, (b) the antenna orientations, and (c) the node orientations. The eye symbols indicate the positions and angles of the pictures shown in \figref{fig:photos}.}
    \label{fig:planPaper6}
    \vspace*{-1mm}
\end{figure}

For this study, a small six-node network was established in a two-story indoor environment as depicted in \figref{fig:planPaper6}.
This setting resembles a typical factory, with nodes spaced 15 to 25\,m apart, surrounded by metal beams, pipes, wooden walls, and various technical equipment.

\subsubsection{Node Deployment}
To replicate typical installation conditions, the nodes are mounted on the surrounding structure without attention to proper antenna alignment.
In the test case show in in \figref{fig:planPaper6}, nodes A and D are mounted onto concrete walls, nodes B, f, and g are mounted vertically onto metal beams, and node C sits on the railing.
Nodes D and f are on the lower floor and the rest are on the upper floor.

\begin{figure}
    \centering
    \subfloat[][]{\includegraphics[width=7.6cm]{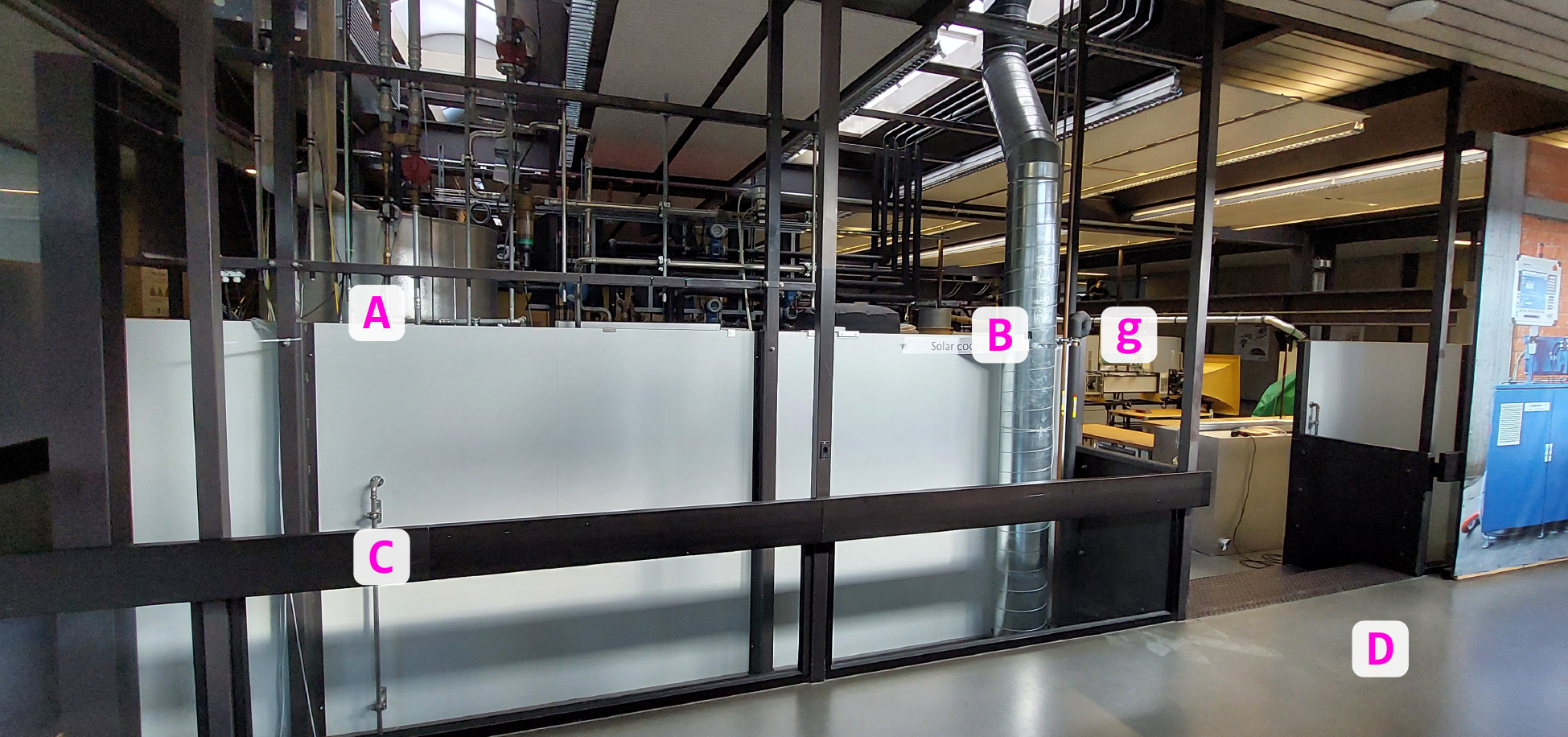}}
    \hfill
    \subfloat[][]{\includegraphics[width=7.6cm]{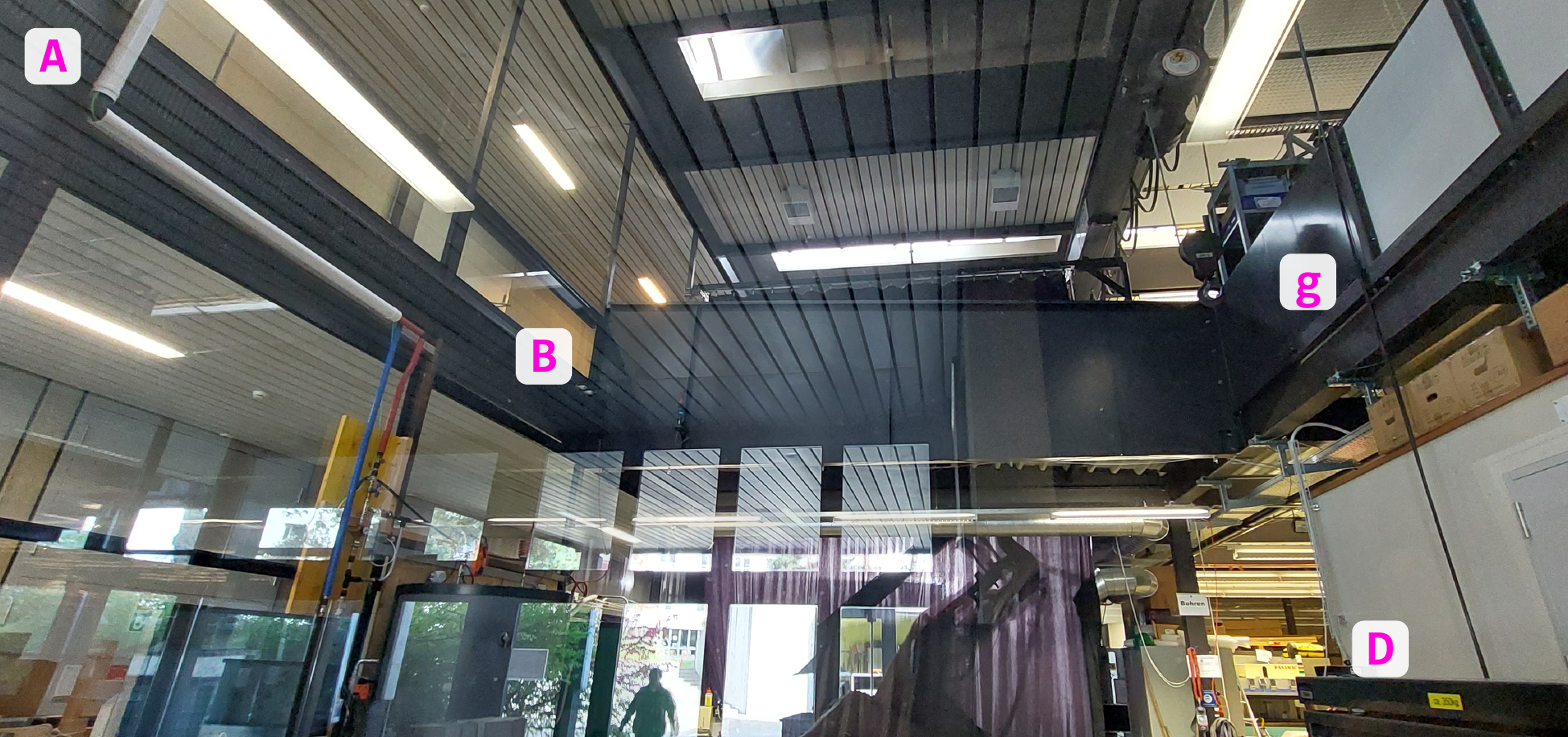}}

    \subfloat[][]{\includegraphics[width=7.6cm]{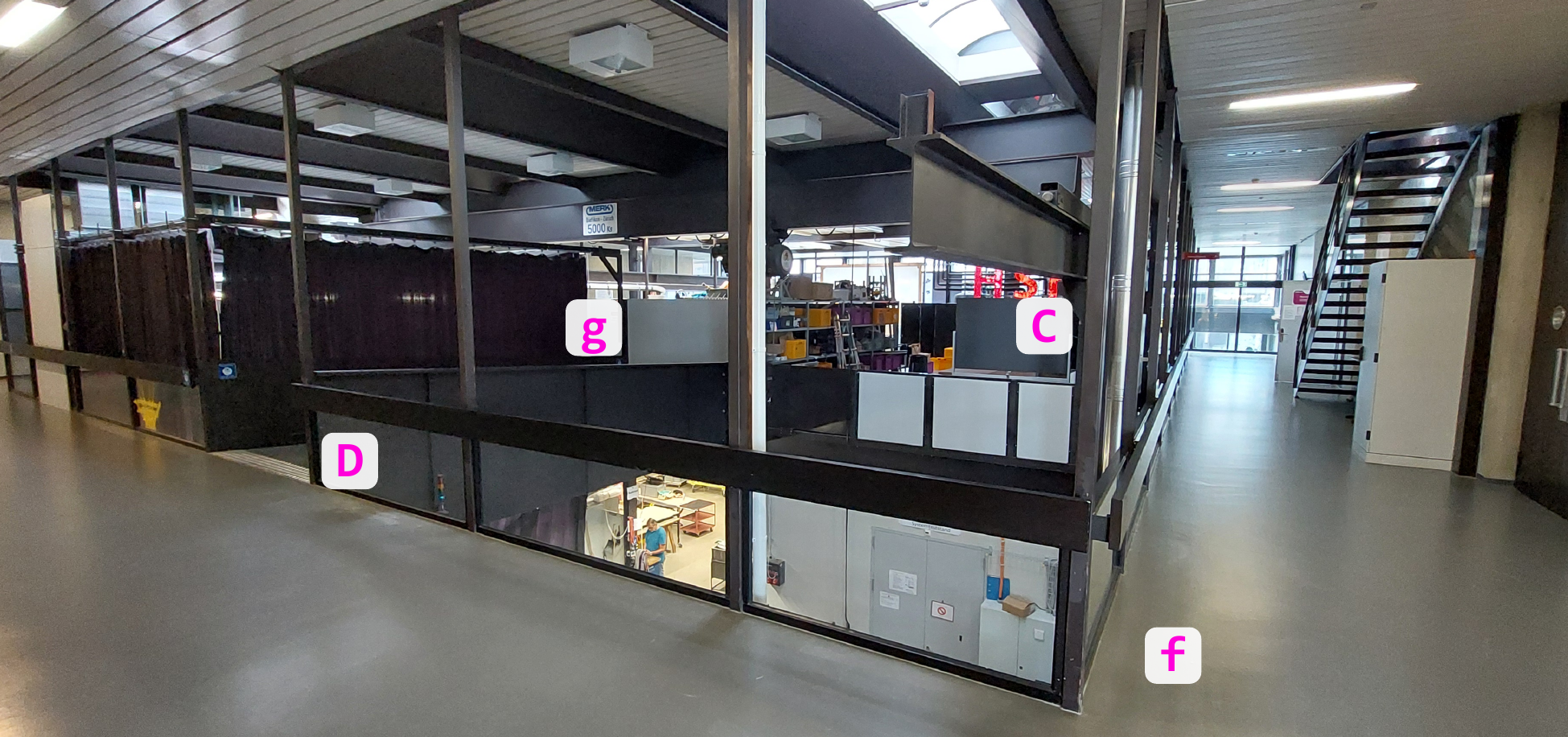}}
    \hfill
    \subfloat[][]{\includegraphics[width=7.6cm]{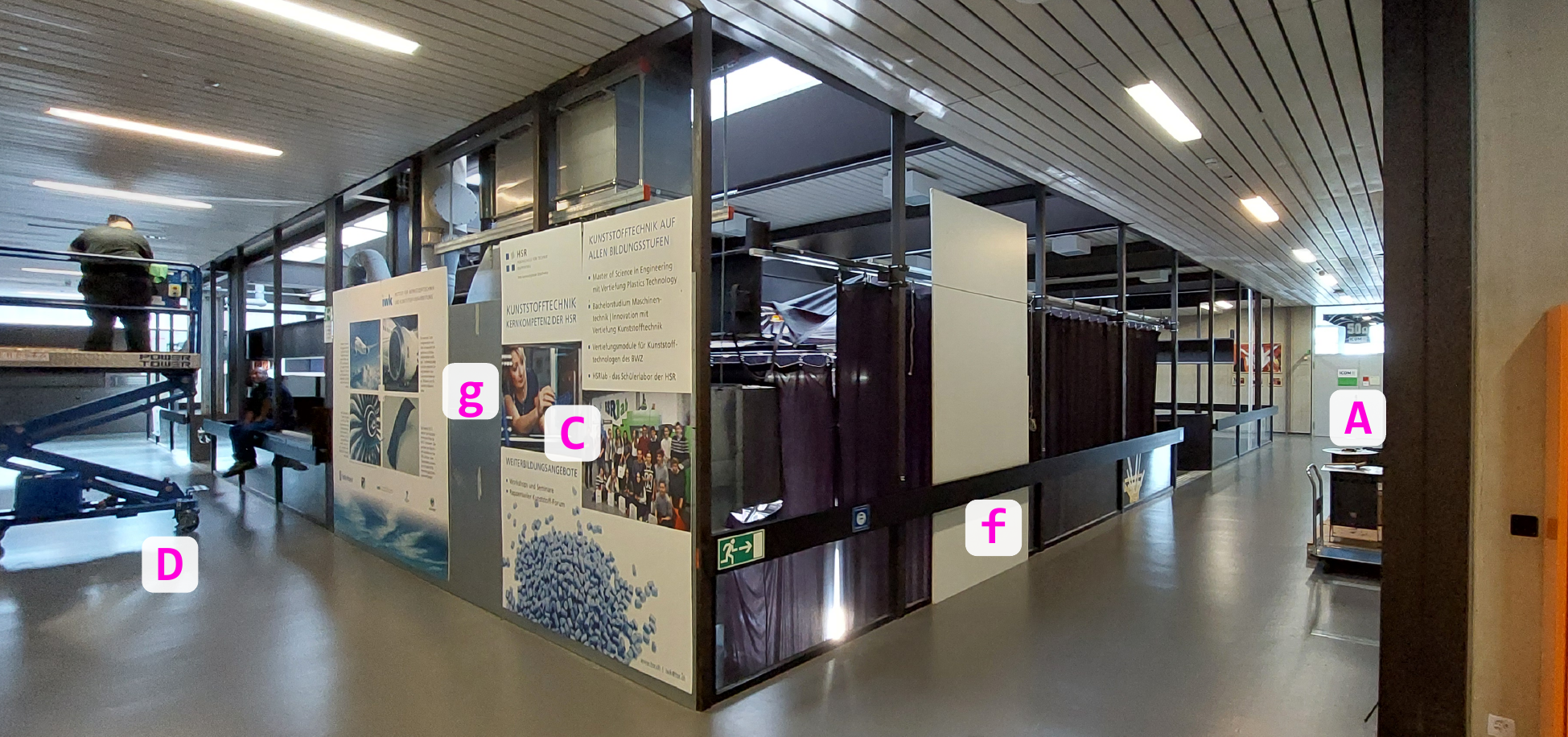}}
    \caption{Views from different node locations into the central structure; the photos (a) -- (d) correspond to the respective eye symbols in \figref{fig:planPaper6}.}
    \label{fig:photos}
    \vspace*{-1.7mm}
\end{figure}

\subsubsection{Test Procedure and Conditions}
The first objective was to measure the Received Signal Strength Indicator (RSSI) for every possible node and antenna combination.
An automated script was used to switch between antennas and collect 10 data points per combination, which were averaged for comparison. To minimize external influences, all tests considered herein were conducted at night.

The second objective was to assess the maximum available throughput using iPerf.
As shown in \figref{fig:throughput_setup}, throughput was measured between two different Raspberry Pis located near the last node, minimizing the effect of the final link between the optimized node and the single-antenna Raspberry Pi.
The throughputs for two different connections (A$\leftrightarrow$g and A$\leftrightarrow$D) were measured, each with four configurations:
\begin{itemize}
    \item All nodes connected via antenna 1 (no diversity).
    \item All nodes connected via antenna 2 (no diversity).
    \item Best PCB antenna for each node (2D antenna diversity).
    \item Best antenna overall for each node (full 3D diversity).
\end{itemize}
This procedure enables a comprehensive evaluation of the full range of antenna configurations and antenna diversity levels, including the practically relevant 2D diversity case using only PCB antennas.

\vspace*{4pt}
\section{Results}

\subsection{Antenna Diversity Gain in Terms of RSSI}
\begin{figure}[!htb]\centering
    \includegraphics[width=8cm,clip,trim={15mm 15.5mm 20mm 23mm}]{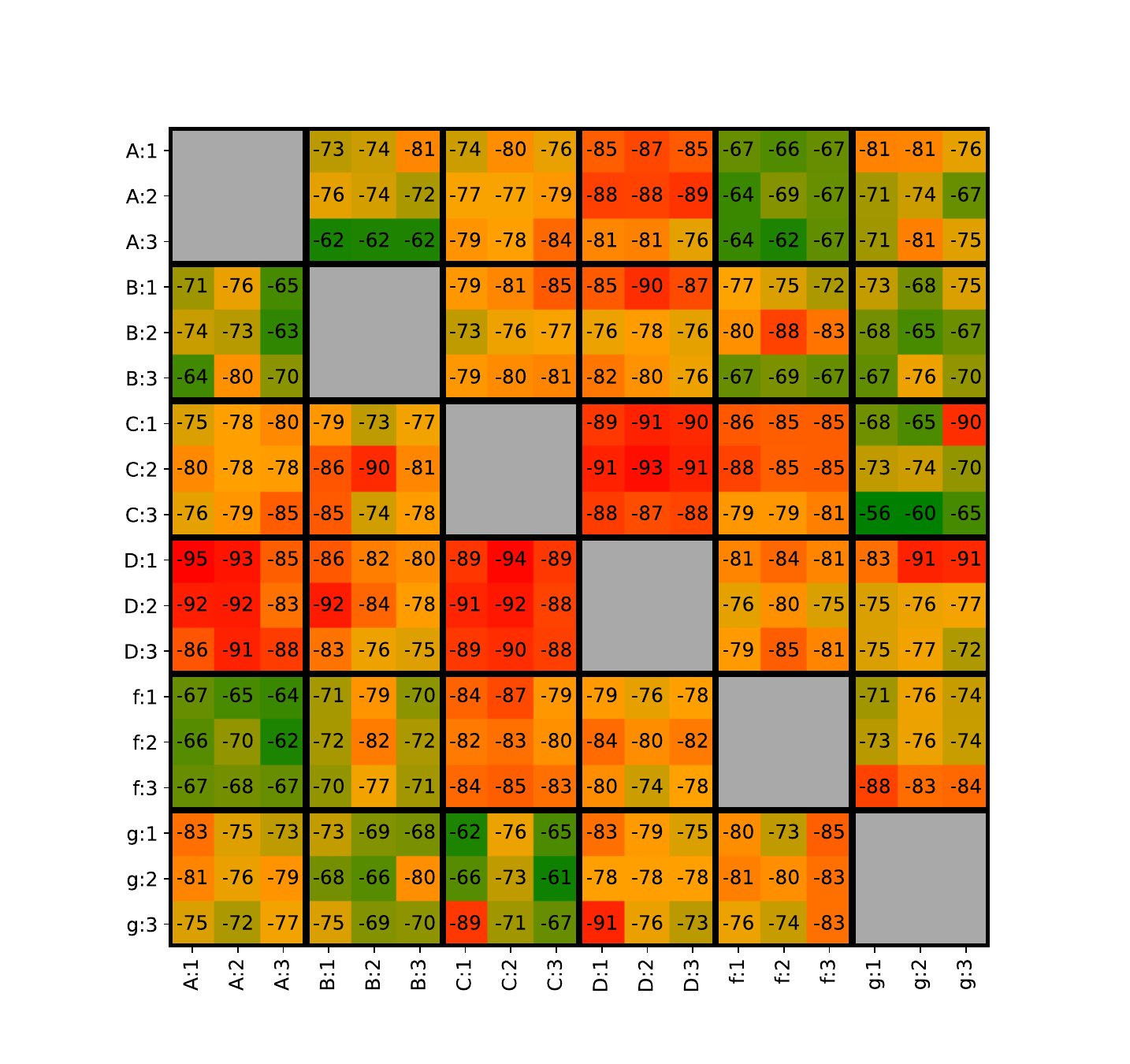}
    \vspace*{-1mm}
    \caption{Full connection matrix with the RSSI values between six nodes, each with three antennas.
        Uppercase identifiers denote large nodes, and lowercase refer to small nodes.
        The number after the column indicates the active antenna, numbered as shown in \figref{fig:planPaper6}(b).
        For the color legend see \figref{fig:matrixPCB}.}
    \label{fig:matrixFull}
\end{figure}

\begin{figure}[!htb]\centering
    \begin{tikzpicture}
        \node[anchor=north west] (img1) at (0, 0) {
            \includegraphics[width=5.33cm,clip,trim={0 7mm 8mm 15mm}]{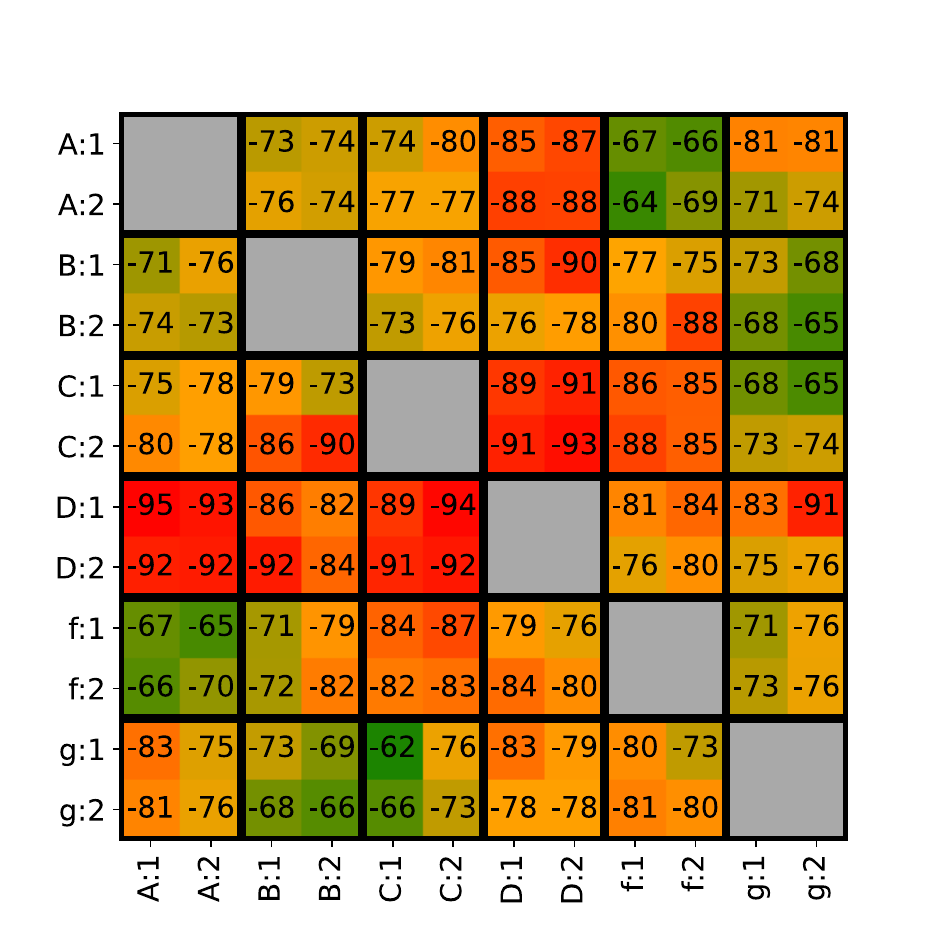}

        };
        \node[anchor=north west, xshift=5.4cm, yshift=-1.31cm] (img2) at (0, 0) {
            \includegraphics[height=3.3cm]{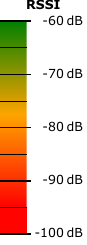}
        };

    \end{tikzpicture}
    \vspace*{-1mm}
    \caption{Subset of \figref{fig:matrixFull}, showing only connections via the two PCB-based antennas (i.e., antenna 1 and antenna 2).}
    \label{fig:matrixPCB}
    \vspace*{-0.5mm}
\end{figure}

The evaluation of antenna diversity enhancement is conducted through matrix representations of the RSSI values, as shown in Figs.~\ref{fig:matrixFull} and \ref{fig:matrixPCB}.
These matrices capture the RSSI readings from each antenna configuration across different nodes and orientations.
By systematically recording and comparing these values, these matrices provide a clear visualization of the signal strength variations attributable to each antenna.

To illustrate the use of these matrices, consider the following example: The link between devices g and A, corresponding to the bottom-left corner of the matrix in Fig.~\ref{fig:matrixFull}, has an RSSI of --83\,dBm when both nodes use antenna 1. Switching device A to antenna 2 improves the link by 8\,dB to --75\,dBm. The best link, however, is obtained when device g is additionally switched to antenna 3, resulting in an RSSI of --72\,dBm\,---\,an improvement of 11\,dB compared to the initial configuration.

This approach highlights the improvements in link quality achieved through antenna diversity, showing how different antenna orientations and configurations could mitigate signal degradation and optimize network performance and reliability.
The matrix representation effectively quantifies  and visualizes the diversity gains in received signal strength.

The results show that antenna diversity provides significant enhancements in received signal strength, with gains ranging from 6 to 34\,dB in this test network.
In some cases, antenna diversity transforms poor or non-existent connections into stable ones, which is especially valuable in environments where reliable network connections are critical, such as industrial or densely populated areas.
The best RSSI improvement of 34\,dB surpasses the 22\,dB difference for depolarized connections reported by Golmohamadi and Frolik \cite{golmohamadi16} and exceeds the improvements achieved by Ramirez et al. \cite{ramirez17,ramirez18} by almost 24\,dB. Such significant improvements can be crucial for IoT applications, greatly impacting performance and reliability.

\begin{figure}\centering
    \includegraphics[width=0.95\columnwidth,clip,trim={0 2mm 0 0}]{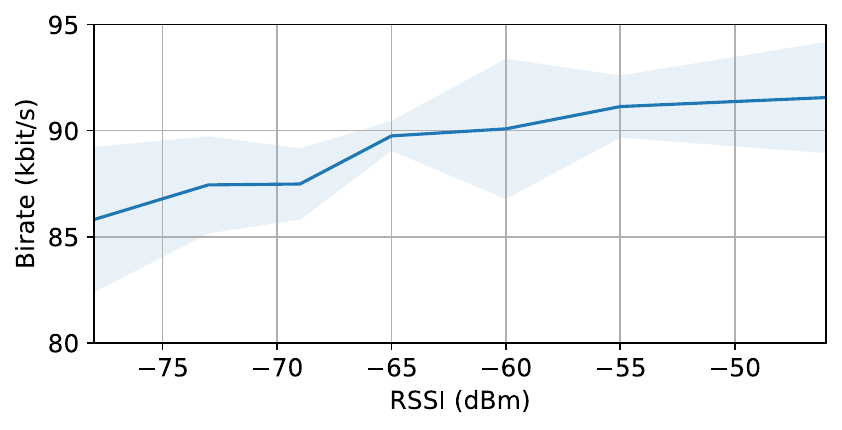}
    \caption{Measured throughput between two Raspberry Pis at varying RSSI values. Each data point represents 100 samples, with a UDP packet size of 512 bytes. The shaded area indicates one standard deviation from the mean value. Note that reliable throughput measurements could not be obtained below \mbox{--78\,dBm} using iPerf.}
    \label{fig:throughput}
\end{figure}

The aforementioned results were all achieved using all three antennas.
However, as mentioned, the third antenna adds significant complexity, so special attention was given to the setup with only two PCB antennas.
Even with just these two antennas, the connections improved by 3 to 17\,dB, highlighting the effectiveness of this diversity level.

\subsection{Antenna Diversity Gain in Terms of Throughput}
In a one-hop scenario as shown in \figref{fig:throughput}, the maximum achievable throughput strongly correlates with the RSSI values.
However, in a multi-hop network, the maximum throughput  highly depends one the number of hops in the connection, as supported by findings from OpenThread \cite{OTquality} and \cite{chowdhury22}.
The measured throughputs in \tabref{tab:throughput} confirm this, showing only small improvements with antenna diversity.
Nevertheless, the connection between node A on the upper floor and node D on the lower floor, spanning at least three hops, increased by 5\%.
These results show that antenna diversity enhances network reliability and robustness rather than throughput capacity.
However, further research must be conducted to assess the reliability of the link by performing a statistical analysis on factors such as packet loss, error rate, and latency.

\begin{table}[h]
    \centering
    \caption{Maximum Bitrates Measured for Two-Node Connections Using Different Antenna Configurations.}
    \begin{tabular}{lcc}
        \toprule
        \bf{Configuration}              & \bf{A $\bm{\leftrightarrow}$ g} & \bf{A $\bm{\leftrightarrow}$ D} \\\midrule
        All nodes use only antenna 1    & 75\,kbit/s                      & 25.5\,kbit/s                    \\
        All nodes use only antenna 2    & 75\,kbit/s                      & 26.0\,kbit/s                    \\
        Choosing the best PCB antenna\; & 78\,kbit/s                      & 27.0\,kbit/s                    \\
        Choosing the best antenna       & 79\,kbit/s                      & 27.0\,kbit/s                    \\
        \bottomrule
    \end{tabular}
    \label{tab:throughput}    \vspace*{-1mm}
\end{table}

\balance
\vspace*{4pt}
\section{Summary \& Conclusions}
This study investigates the application of antenna diversity to enhance the reliability, robustness, and performance of Internet of Things (IoT) networks through measurements conducted in indoor environments with real-world conditions.
The integration of multiple antennas with different polarizations addresses common challenges such as antenna misalignment and interference from surrounding objects, which can significantly degrade received signal quality and network reliability.

The experiments yielded diversity gains in received signal strength of up to 34\,dB for individual connections, with a minimum gain of 6\,dB for each single connection, compared to the worst-performing antenna configuration.
Even with the simpler, more practical, and cost-effective solution of implementing only two PCB antennas, diversity gains between 3 to 17\,dB were observed.
While antenna diversity primarily improves reliability and robustness rather than throughput, a 5\% increase in data throughput was also observed. These improvements are critical in challenging environments, where stable and robust connections are essential.

Overall, these experiments confirmed that antenna diversity can effectively mitigate signal degradation and significantly enhance network reliability and robustness.
While incorporating multiple antennas and the associated RF components increases the complexity of hardware and software design,
the benefits gained in terms of improved signal quality and network reliability likely justify the additional effort.
These findings suggest that antenna diversity solutions may well be worth the investment for demanding IoT applications, particularly in dynamic and challenging environments, such as those typical of manufacturing, healthcare, and logistics.

\end{document}